\documentclass[10pt,prl,aps,showpacs,twocolumn,unsortedaddress]{revtex4-1}
\usepackage{graphicx,bm}
\usepackage{amsmath}
\usepackage{amssymb}
\usepackage{epsfig}
\usepackage{verbatim}
\usepackage{hyperref}
\usepackage[usenames]{color}

\usepackage{pdfpages}

\newcommand{\dd}{\, \mathrm{d}}

\newcommand{\ve}{\varepsilon}

\newcommand{\bs}{\boldsymbol}

\def\be#1\ee{\begin{equation}#1\end{equation}}
\def\ba#1\ea{\begin{align}#1\end{align}}
\def\bal#1\eal{\begin{equation}#1\end{equation}}

\begin{document}

\title{
Quantum rotor model for a Bose-Einstein condensate of dipolar molecules
}

\author{J. Armaitis}
\email{j.armaitis@uu.nl}
\author{R.A. Duine}
\author{H.T.C. Stoof}
\affiliation{Institute for Theoretical Physics, Utrecht
University, Leuvenlaan 4, 3584 CE Utrecht, The Netherlands}

\date{\today}

\begin{abstract}
We show that a Bose-Einstein condensate of heteronuclear molecules in the regime
of small and static electric fields is described by a quantum rotor model for the macroscopic electric dipole
moment of the molecular gas cloud.
We solve this model exactly and find the symmetric, i.e., rotationally invariant, and dipolar phases
expected from the single-molecule problem, but also an axial and planar nematic phase due to
many-body effects.
Investigation of the wavefunction of the macroscopic dipole moment also reveals squeezing of the
probability distribution for the angular momentum of the molecules.
\end{abstract}

\maketitle

\textit{Introduction.---} 
A promising new direction in the field of ultracold quantum gases is the study
of dipolar gases with heteronuclear molecules
\cite{MoleculeSengstock,MoleculeGroundstateJin,MoleculeStableZwierlein}.
Recent progress in this direction has already contributed to 
such diverse research areas as atomic and molecular physics, quantum
computation, and chemistry \cite{Review1,Review2,krems2010cold}.
Indeed, the unique combination of strongly anisotropic long-range interactions and the quantum nature in these systems
has brought to light a number of striking phenomena, such as tunneling-driven \cite{UltracoldTunneling} and direction-dependent \cite{directionchemistry} ultracold chemical reactions, 
as well as the shape-dependent stability of the gas cloud \cite{ShapeStability}.

The novel ingredient of heteronuclear molecules as compared to neutral atoms is their large
permanent electric dipole moment, which opens the possibility for a strong dipole-dipole interaction.
Neutral atoms typically do have a permanent magnetic dipole moment,
but this leads to a dipole-dipole interaction that is much weaker than in the case of
heteronuclear molecules, although it nevertheless has observable effects in certain
cases \cite{RydbergInBEC,AtomicDy}, in particular when the scattering length is made 
small using a Feshbach resonance \cite{AtomicCr,HuletSmallSwave,AtomicEr}.
In the absence of an external electric field, however, the average dipole moment in the 
laboratory frame is zero, since the rotational ground state of the molecule is spherically symmetric
and the dipole moment is thus randomly oriented. For that reason, virtually all theoretical 
many-body studies are carried out in the limit of a large DC electric field. In that limit the molecules 
are completely polarized and the dipole moment in the laboratory frame is maximal \cite{ZollerChapter}.
One notable deviation from the large electric field limit is the discussion by Lin \emph{et al.} \cite{Ferroelectricity}, which considers the effects of an almost resonant AC electric field.

\begin{figure}[t]
\begin{center}
\includegraphics[width=0.95\linewidth]{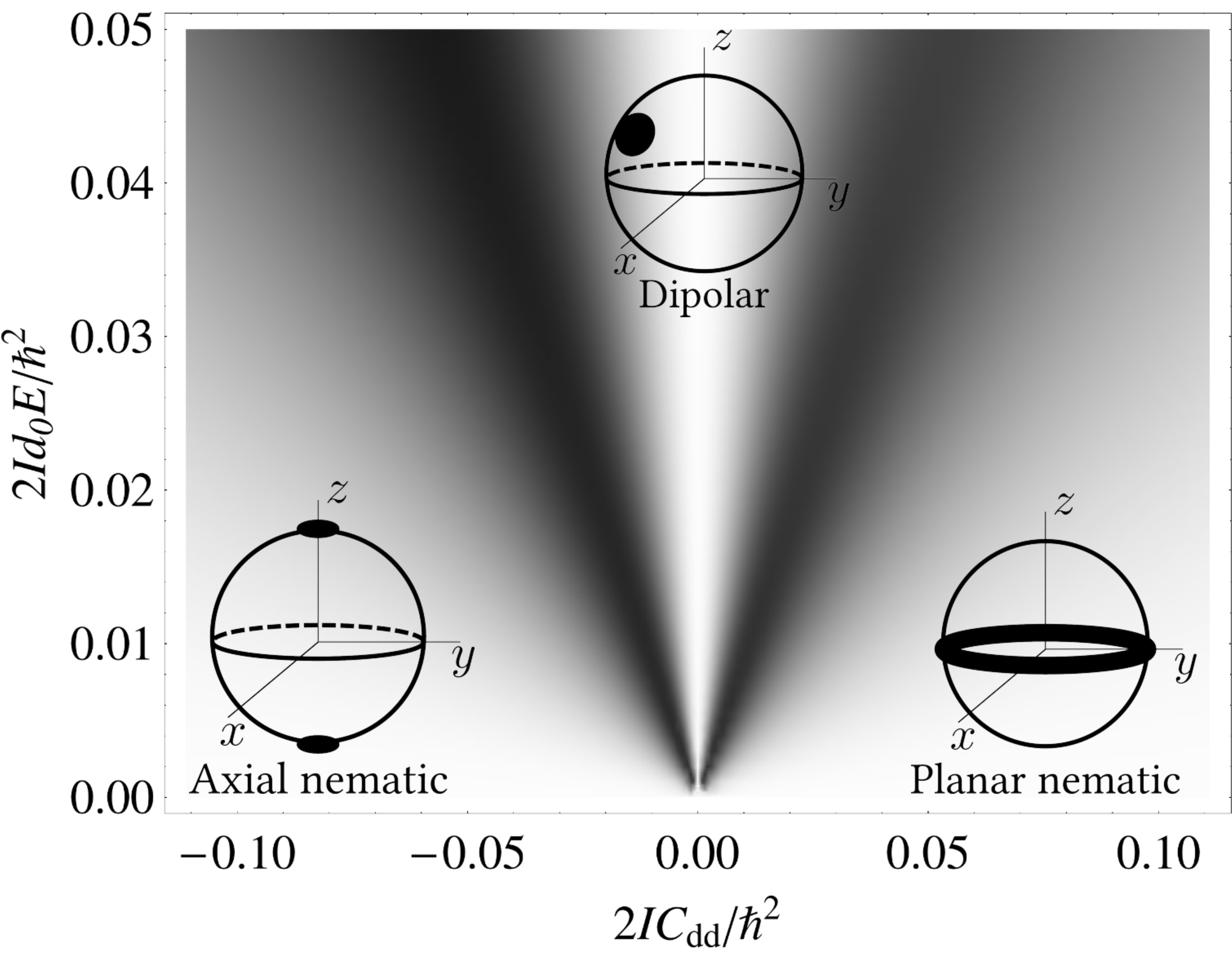}
\caption{
Phase diagram of the axially symmetric Bose-Einstein condensate of
heteronuclear molecules, where the probability distributions for the dipole
moment on the unit sphere in the non-trivial
phases are schematically indicated by the black areas on the spheres.  The
vertical axis is the external electric field, while the horizontal axis is the
dipole-dipole interaction strength. In this diagram, the fully symmetric phase
exists only in the origin.  Shading corresponds to the ``squeezing" parameter
$\sigma$ in Eq. \eqref{squeezing-parameter}, which runs from zero (white) to $0.09$ (gray).  The electric
field is at an $\pi /4$ angle to the symmetry axis of the cloud.
}
\label{fig:pd}
\end{center}
\end{figure}

Going away from the large-field limit unmasks the subtle interplay between the
quantum-mechanical rotation of the molecules, the long-range dipole-dipole
interaction and the directing static electric field, which is the main topic of this
Letter.  In particular, the molecular Bose-Einstein condensate turns
out to be a ferroelectric material that is fully disordered by quantum
fluctuations in the absence of an electric field. This is illustrated by
the phase diagram of a Bose-Einstein
condensate of heteronuclear molecules in a harmonic uniaxial trap, that is shown in Fig. \ref{fig:pd}. The
system possesses four phases: two nematic phases (a planar nematic and an axial
nematic phase), a dipolar phase, and a fully symmetric phase, that are separated by
smooth crossovers.  
Two order parameters are relevant for this system. Firstly, a non-zero average dipole moment
$\langle d_i \rangle$ defines the dipolar phase. 
Secondly, in the absence of an average dipole moment the nematic (or quadrupole) tensor $Q_{ij} = \langle d_i d_j  - \delta_{ij} \bs d^2 /3\rangle$
distinguishes the other three phases. In particular, the nematic tensor is equal to zero in the spherically
symmetric phase. Two eigenvalues are positive and one is negative in the planar nematic
phase, whereas one eigenvalue is positive and two are negative in the axial nematic phase.
It is worthwhile to notice that even in the absence of any
electric field, many-body effects are crucial, giving rise to nematic ground
states in strong contrast to the dipolar and fully symmetric ground states,
expected from the single-molecule case.  We finally remark that the predicted
phase diagram is experimentally accessible by tuning three parameters in the
laboratory, namely, the electric-field strength, the trap aspect ratio, and the number of
particles.

\textit{Model.---} We start from the single-molecule Hamiltonian
\be
\label{single-molecule-hamiltonian}
H_\text{m} = 
\frac{\bs p^2}{2m} +
\frac{\bs L^2}{2I} - d_0 \boldsymbol{\hat d} \cdot \bs E,
\ee
where $m$ is the mass of the molecule, $\bs p = -i \hbar \, \partial / \partial \boldsymbol x$ is the 
center-of-mass momentum operator with $\bs x$ the center-of-mass position, 
$I$ is the moment of inertia of the molecule, 
$d_0 \boldsymbol{\hat d}$ is the electric dipole moment operator,
$\bs L = -i \hbar \boldsymbol{\hat d} \times \partial / \partial \boldsymbol{\hat d}$ 
is the angular momentum operator, associated with the rotation of the molecules, and $\bs E$ is the electric field. To describe the interactions between the molecules, we have to include both
a contact (or $s$-wave) interaction term \cite{stoofbook}
\be
V_\text{s} = \frac{4\pi \hbar^2 a}{m}\delta(\bs r),
\ee
and a dipole-dipole interaction term
\be
V_\text{dd} = - \frac{d_0^2}{4 \pi \ve_0 r^3}
\left(
3 \, \boldsymbol{\hat d}_1 \cdot \boldsymbol{\hat r} \,  \boldsymbol{\hat d}_2 \cdot \boldsymbol{\hat r}
- \boldsymbol{\hat d}_1 \cdot \boldsymbol{\hat d}_2
\right),
\ee
where $\delta$ is the Dirac delta function, $a$ is the $s$-wave scattering length, $\ve_0$ is the electric permittivity of vacuum, $d_0 \boldsymbol{\hat d}_1$ and $d_0 \boldsymbol{\hat d}_2$ are the dipole moments of the two interacting particles, $\boldsymbol{r}$ is the vector
connecting them and $r$ is the distance between the particles. 
Finally, we consider the molecular gas to be trapped in
a harmonic axially-symmetric trapping potential
\be
V_\text{trap} = m \left[ \omega_\perp^2 (x^2 + y^2) + \omega_z^2 z^2\right]/2,
\ee
where $\omega_\perp$ and $\omega_z$ are the radial and axial trapping frequencies, respectively.

For small electric fields, we are allowed to first solve for the spatial
part of the condensate wavefunction by only
including the effect of the $s$-wave interaction between the molecules.
This leads to a Thomas-Fermi profile that depends on the $s$-wave scattering length \cite{Pethick:1136200}. The many-body ground state wavefunction is now
$
\Psi(\bs r_1, \bs r_2, \ldots, \bs r_N; \bs d_1, \bs d_2, \ldots, \bs d_N)
= \Pi_{i=1}^N \psi_\text{TF}(\bs r_i) \times \psi(\boldsymbol{\hat d}),
$
where $N$ is the total number of molecules, and $\boldsymbol{\hat d}$ is the direction of $\bs d=(\sum_{i=1}^N \bs d_i)/N$.
Hence, the dipole-dipole energy per particle is
\be
V^\text{TF}_\text{dd}
= 
- \frac{N d_0^2}{4 \pi \ve_0}
\int \dd \bs r
P(\bs r)
\frac{1}{r^5}
\left(
3 \, (\boldsymbol{\hat d} \cdot \bs r)^2
- \boldsymbol{\hat d}\,^2  r^2
\right),
\ee
where  
$P$ is the probability to find two particles a certain distance apart.
Subsequently, the many-body Hamiltonian per molecule in this so-called single-mode approximation \cite{PhysRevLett.81.5257} reduces to (c.f. Sec. I of Ref. \cite{supp})
\be
\label{many-body-H}
H
=
\frac{\bs L^2}{2I} 
- d_0 \boldsymbol{\hat d} \cdot \bs E
+ C_\text{dd} (3 {\hat d}_z^2 - \boldsymbol{\hat d}\,^2),
\ee
where $C_\text{dd}$ is the effective dipolar interaction strength
\be
C_\text{dd} = \frac{d^2_0 N}{4 \ve_0}  \int \dd z \rho \dd \rho 
P(R) \frac{1}{r^3}
\left( \frac{3}{2} \frac{\rho^2}{r^2} - 1 \right),
\ee
and we have introduced the radius in cylindrical coordinates $r^2 = \rho^2 + z^2$, the dimensionless
radius $R^2 = (\rho/x_\text{TF})^2+(z/z_\text{TF})^2$,
the radial size of the cloud $x_\text{TF}$, the axial size $z_\text{TF} = \lambda x_\text{TF}$, 
and the aspect ratio $\lambda = \omega_\perp / \omega_z$.
We emphasize that even though Eq. \eqref{many-body-H} describes the
dipole degree of freedom of the
three-dimensional many-body system, it actually has a form of a single-particle (and thus effectively zero-dimensional) Hamiltonian, and the whole Bose-Einstein condensate acts as a single quantum
rotor.

In the Thomas-Fermi approximation the probability $P$ can be calculated analytically:
$P(R) = 15 (R-2)^4 \left(32+64 R+24 R^2+3 R^3\right)/7168 \pi \lambda x_\text{TF}^3$ 
for $R<2$ and zero otherwise. The analytic expression for $P(R)$ yields \cite{PhysRevLett.89.130401,PhysRevA.74.013621}
\ba
\label{aa}
&C_\text{dd} = -5Nd_0^2/\left({56 \pi \ve_0 x_\text{TF}^3 \lambda  \left(\lambda ^2-1\right)^2}\right)
\\
\nonumber
&\times 
\left(\lambda ^4+\lambda ^2-2+3 \lambda  \sqrt{1-\lambda ^2} \text{ArcCot}\left[\frac{\lambda }{\sqrt{1-\lambda ^2}}\right]\right),
\ea
which corresponds to one half of the mean-field dipolar energy per particle in the case of fully polarized electric dipoles (\cite{PhysRevLett.89.130401,PhysRevA.74.013621}).
Analogous results for magnetic dipoles were obtained by other authors for spinor Bose-Einstein
condensates in the Gaussian approximation \cite{GaussianCdd,PhysRevLett.93.040403}.
Note that $C_\text{dd}$ depends on the number of particles $N$ and the
trap aspect ratio $\lambda$. Thus, the only effect of varying the number of particles
is the change in $C_\text{dd}$.

The Hamiltonian in Eq. \eqref{many-body-H} represents a quantum rotor model for the
macroscopic dipole moment of the molecular Bose-Einstein condensate, whose derivation
is the main result of this Letter. 
Interestingly, a similar Hamiltonian applies to an atomic ferromagnetic spinor
Bose-Einstein condensate 
(c.f. Ref. \cite{AntiferRotor} for a quantum rotor model of antiferromagnetic spinor condensates), 
but then without the quantum rotor term \cite{GaussianCdd}. The
reason for this difference is that the total (spin) angular momentum of the
atoms is fixed, whereas in the case of interest here the wavefunction of the
molecules is in general a superposition of states with an arbitrary (rotational)
angular momentum, whose energy splitting is determined by the finite moment of
inertia. Next we are going to investigate the ground-state properties
of this quantum rotor model.

\textit{Results.---} 
We have obtained the exact phase diagram pertaining to this Hamiltonian by expanding
the dipolar wavefunction in spherical harmonics (Fig. \ref{fig:pd}). For zero electric field and no dipole-dipole interaction,
the ground state of the system is a trivial spherically symmetric (non-dipolar) state. 
However, turning on $\bs E$ or $C_\text{dd}$ results in
a very different state. For zero $C_\text{dd}$ and non-zero $\bs E$, we obtain a dipolar state, where the probability
distribution on the sphere is concentrated around the direction of the electric field. This state is classical
in the sense that it is analogous to a classical dipole in the electric field.
Another limiting case is $\bs E=\bs 0$ and $C_\text{dd}<0$, where we have an axial nematic phase, and the probability
is concentrated around the north and south poles of the sphere. Finally, we have a planar nematic phase
for $\bs E=\bs 0$, $C_\text{dd}>0$, where the high probability region is located around the equator of the sphere. The last
two phases are quantum mechanical, as the ground state there is a coherent 
superposition of spherical harmonics with no average dipole moment. 
We observe smooth crossovers between the non-trivial phases, as expected due to the 
existence of quantum fluctuations in this effectively zero-dimensional situation.

\begin{figure}[t]
\begin{center}
\includegraphics[width=0.95\linewidth]{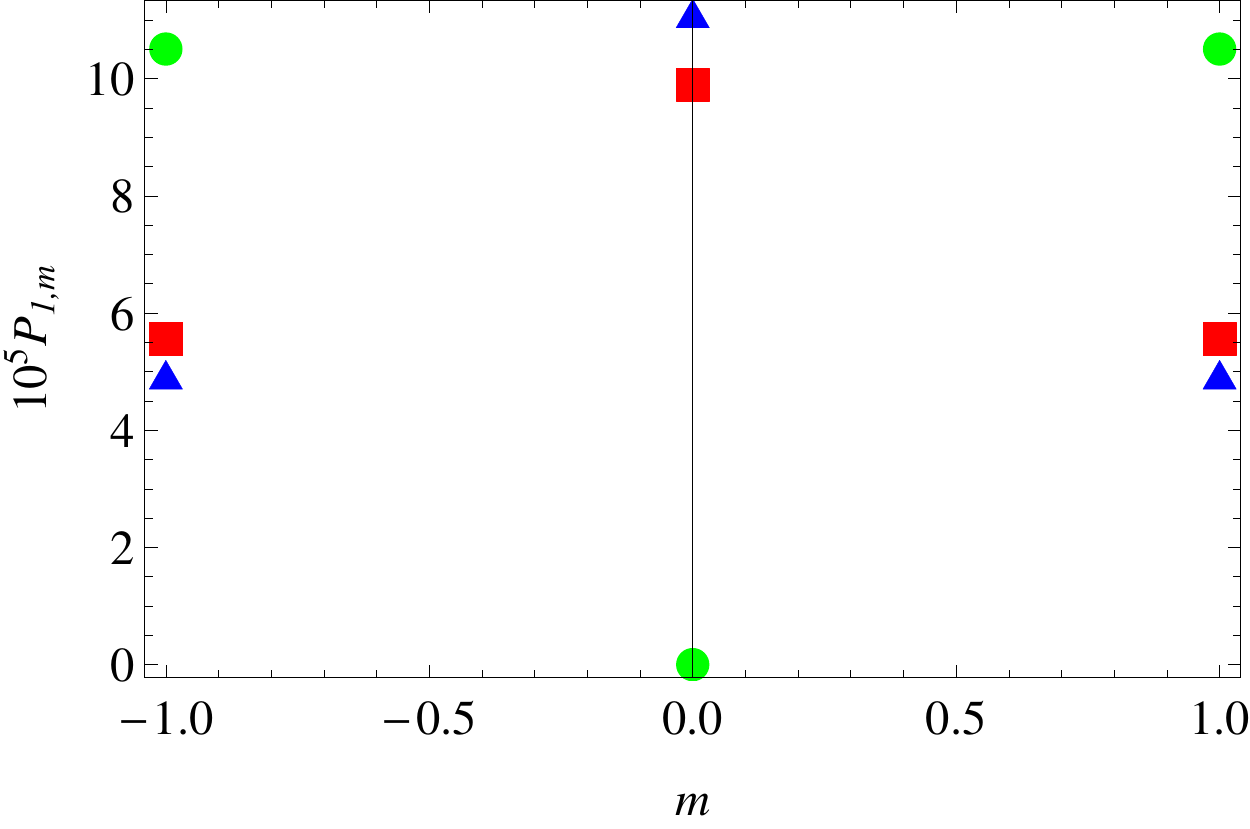}
\caption{Probability $P_{1,m}$ of occupying a state with total angular momentum $1$ and its projection $m$.
We have chosen $C_\text{dd} = 0.1 \hbar^2/2I$ and $E=0.05 \hbar^2/2Id_0$ ($\bs E$ is at $\pi/4$ angle to the $z$ axis) in order to maximize the anisotropy of the state. The red squares correspond to the $x'$ direction,
the green circles correspond to the $y'$ direction, and the blue triangles correspond to the $z'$ direction,
where the axes are defined such that the $\langle \hat d_i \hat d_j\rangle$ matrix is diagonal and has its smallest
eigenvalue in the $z'$ direction.
}
\label{fig:m-probability}
\end{center}
\end{figure}

In addition to the coordinate-space probability distribution $\vert \psi(\boldsymbol{\hat d}) \vert^2$, 
we investigate the probability 
distribution with respect to angular momentum $P_l$. To that end, we expand our wavefunction
in terms of spherical harmonics: $\psi(\boldsymbol{\hat d}) = \sum_{l,m} \alpha_{l,m} Y_{l,m}(\boldsymbol{\hat d})$. Hence, $P_l = \sum_{m=-l}^l P_{l,m}$, where $P_{l,m} = \vert \alpha_{l,m} \vert^2$ is the
probability to occupy a state which has angular momentum quantum number $l$ and 
azimuthal quantum number $m$. 
We find that this distribution has a peak at $l = 0$ for negative $C_\text{dd}$, and is peaked at $l\geq 0$ for
positive $C_\text{dd}$ or non-zero $\bs E$. For larger values of $C_\text{dd}$ and $\vert E \vert$, the peak shifts towards
larger values of $l$. Moreover, due to the nature of the dipole-dipole interaction that conserves
parity, at zero electric
field $P_l$ is zero for odd $l$. 
We have also investigated the distribution of probability between different $\vert l, m \rangle$ states
(Fig. \ref{fig:m-probability}). In general, this distribution is symmetric ($P_{l,m} = P_{l,-m}$) in every 
direction, implying that the average angular momentum $\langle \bs L \rangle$ is always zero, which is a 
consequence of time-reversal symmetry.

Noticing an anisotropic distribution of average dipole moment probability on the sphere in our system
for certain parameters, it is 
natural to draw a parallel with the
effect of spin squeezing \cite{SpinSqueezing1}. To that end, we define a matrix $\langle L_i L_j \rangle$.
This matrix describes the (Heisenberg) uncertainty in the angular momentum of the system. 
It has three eigenvalues, that we order as follows: $|\lambda_0| \leq |\lambda_-| \leq |\lambda_+|$.
Hence, we define a measure of angular momentum ``squeezing" as
\be
\label{squeezing-parameter}
\sigma = \frac{|\lambda_+|-|\lambda_-|}{|\lambda_+|+|\lambda_-|},
\ee
which tells us how anisotropic the uncertainty of angular momentum is (c.f. Fig. \ref{fig:pd}). However,
we must point out that, strictly speaking, this effect is not identical to squeezing in the usual sense, because 
$\langle L_i \rangle=0$ and $P_{l,m}$ is not
always a monotonically decreasing function of $m$ (as can be seen from Fig. \ref{fig:m-probability}).

\textit{Discussion and conclusion.---}  
It is interesting to compare the exact results described so far with mean-field theory techniques
commonly employed for atomic Bose-Einstein condensates.
Thus we turn to the Hartree approximation (which is equivalent to solving the Gross-Pitaevskii equation) for an analysis of the Hamiltonian in Eq. \eqref{many-body-H}. To that end, we replace the operator 
${\hat d}_i^2$ by ${\hat d}_i \langle {\hat d}_i \rangle$. 
The effect of the dipole-dipole interaction is then an additional static electric field of the form
\be
\bs E_\text{dd} = \frac{C_\text{dd}}{d_0}
( \langle \hat d_x \rangle, \langle \hat d_y \rangle, -2 \langle \hat d_z \rangle )^T,
\ee
where the angle brackets indicate a quantum-mechanical average,
and $N \langle \bs d \rangle \cdot \bs E_\text{dd}$ is the total average (Hartree) energy of all the
classical dipoles with a density distribution given by the Thomas-Fermi profile.
Therefore, we now have to solve the effective single-particle Hamiltonian
\be
\label{mean-field-hamiltonian}
H_\text{MF} = \frac{\bs L^2}{2I} - d_0 \boldsymbol{\hat d} \cdot \bs E_\text{eff},
\ee
where $\bs E_\text{eff} = \bs E + \bs E_\text{dd}$ is the effective electric field, which now depends on the
cloud geometry and the average dipole moment.

The average dipole moment in this approach is determined in two steps. First, 
we calculate
the average dipole moment 
of the ground state 
$\langle \bs d\rangle (\bs E )$
from Eq. \eqref{mean-field-hamiltonian}
(see e.g.\ Ref.\ \cite{Review1}). Second, we write down a self-consistency condition, accounting for the
effective electric field:
\be
\label{self-consistent-d}
\langle \bs d\rangle = \langle \bs d\rangle ( \bs E_\text{eff} (\langle \bs d\rangle) ).
\ee
In the well-known case of a single molecule, $C_\text{dd}$ is zero, Eq. \eqref{self-consistent-d} has a single solution, and $\langle \bs d \rangle$ always points in the direction of $\bs E$. However, this is not the case for the
whole $(C_\text{dd}, \bs E)$ space and therefore requires a more thorough analysis. 
For small non-zero $\vert C_\text{dd} \vert$ and $\bs E = 0$, there still is only one solution,
namely, $\langle \bs d\rangle = 0$.
For $C_\text{dd} < 0$ and $E_z \neq 0$, the average dipole moment is always non-zero (Eq. 
\eqref{self-consistent-d} has a single solution), as 
then we are dealing with an Ising-like (easy-axis) model, and $E_z$ couples directly to the
order parameter $\langle \bs d\rangle$. In contrast to this, three solutions exist for $E_z = 0$
and $C_\text{dd}/|E_\perp|$ sufficiently large and negative. The two
$\langle d_z \rangle \neq 0$ solutions are degenerate in energy, while the $\langle d_z \rangle = 0$ 
solution has a higher energy.
On the other hand, for $C_\text{dd} > 0$ we are dealing with an XY-like (easy-plane)
model and thus in that case one obtains a similar non-trivial situation for $E_\perp = 0$ and 
$C_\text{dd}/|E_z|$ large and positive.

When comparing the mean-field theory with the exact diagonalization of the Hamiltonian in 
Eq. \eqref{many-body-H}, it is important to notice that the mean-field ansatz explicitly assumes
that the average dipole moment is pointing in some direction. Therefore, the nematic phases are absent from the 
mean-field phase diagram.
The fact that the dipole moment is zero under
the asserted conditions can be intuitively understood, as the exact approach allows for a quantum superposition
of states that have oppositely polarized dipole moments and are degenerate at the mean-field level. Finally, it is well known that
the mean-field theory does not give reliable results in low dimensions because of the increased importance
of quantum fluctuations. Since we are investigating an effectively zero-dimensional Hamiltonian, it is no surprise
that the results of the mean-field theory differ significantly from the exact calculation.

In our analysis we have relied on the single-mode approximation, which is applicable
to Bose-Einstein condensates with $s$-wave and dipole-dipole interactions \cite{TFworksForDipoles}. 
However, we have not accounted for the dependence of the cloud aspect ratio $z_\text{TF}/x_\text{TF}$ 
on dipole-dipole interactions. This limits the applicability of our analysis to the regime, where 
the dipole-dipole interaction is much weaker than the mean-field $s$-wave interaction \cite{GaussianCdd}, i.e.,
$\vert\langle \bs d \rangle\vert^2 m / 4\pi \hbar^2 \ve_0 a \ll 1$. For a typical diatomic
molecule with a mass of 80 atomic mass units, a scattering length of $5$ Bohr radii and an electric dipole moment of $1$ Debye,
this limits the external electric field strength to $\vert \bs E \vert \ll 1 \, \text{kV}/\text{cm}$, which translates to $2Id_0 E / \hbar^2 \ll 0.05$ in the units of Fig. \ref{fig:pd}. We have also estimated that
for a cloud of $10^7$ particles with a linear extent of around $1\mu m$, or
radial trapping frequency of approximately $2\pi \times 80$ kHz, assuming a nearly two-dimensional trap with an aspect ratio of 1:10, $C_\text{dd} \simeq 0.1 \times \hbar^2/2I$, which corresponds to the energy of $2\pi \hbar \times 1$ GHz.

Besides the single-mode approximation, we have also made an assumption that the $s$-wave
scattering length is independent of the dipole moment. Even though
it has been shown that such a dependence is present \cite{AonD1,AonD2,AonD3,AonD4},
including it would merely add an extra self-consistency equation to our approach. Its effect would be to change the Thomas-Fermi radii and thus map the system to a different point in the phase diagram. Therefore, all our results
remain qualitatively unaffected.

In summary, we have considered an interacting Bose-Einstein condensate
of dipolar molecules in a small static electric field. We have solved 
this problem exactly in the single-mode approximation and
have also compared this with the mean-field (Gross-Pitaevskii)
approach. We have found that the two approaches to the 
problem yield qualitatively very different results. Finally, we have put forward an 
experimentally accessible phase diagram and investigated the exact ground-state wavefunction both in 
coordinate and angular-momentum space.

This work is supported by the Stichting voor Fundamenteel Onderzoek der Materie (FOM) and the Nederlandse Organisatie voor Wetenschaplijk Onderzoek (NWO).

\bibliography{dipole-prl}

\onecolumngrid
\newpage
\newpage



\section*{Supplementary material}

# Supplemental material for "Quantum rotor model for a Bose-Einstein condensate of dipolar molecules"

J. Armaitis,[*] R. A. Duine, and H. T. C. Stoof
*Institute for Theoretical Physics, Utrecht University,*
*Leuvenlaan 4, 3584 CE Utrecht, The Netherlands*
(Dated: November 4, 2013)

This supplemental material file provides a detailed derivation of the single-mode approximation (SMA) Hamiltonian and also a comment on the probability for the macroscopic dipole moment to occupy a state with total angular momentum $l$ and its projection $m$.

## I. SINGLE-MODE APPROXIMATION HAMILTONIAN

In this section we provide a more detailed derivation of the SMA Hamiltonian [1] than the one given in the Letter. We start from a many-body Hamiltonian with interactions

$$H_{\rm mb} = \sum_{i=1}^{N} \left( \frac{\bm{p}_i^2}{2m} + \frac{\bm{L}_i^2}{2I} - d_0 \hat{\bm{d}}_i \cdot \bm{E} + V_{\rm trap}(\bm{r}_i) \right) + \sum_{i<j} \left( V_s(\bm{r}_i - \bm{r}_j) + V_{dd}(\hat{\bm{d}}_i, \hat{\bm{d}}_j, \bm{r}_i - \bm{r}_j) \right), \quad (1)$$

where the s-wave interaction is

$$V_s(\bm{r}) = \frac{4\pi\hbar^2 a}{m} \delta(\bm{r}), \quad (2)$$

the trapping potential is

$$V_{\rm trap}(\bm{r}) = m \left[ \omega_\perp^2 (x^2 + y^2) + \omega_z^2 z^2 \right]/2, \quad (3)$$

and where $\bm{r}_i$ is the center-of-mass position of molecule $i$, $d_0 \hat{\bm{d}}_i$ is its electric dipole moment ($\hat{\bm{d}}_i$ is the direction of $\bm{d}_i$), and

$$\bm{L}_i = -i\hbar \hat{\bm{d}}_i \times \partial/\partial \hat{\bm{d}}_i \quad (4)$$

is its angular momentum. Moreover, the dipole-dipole interaction is

$$V_{dd}(\hat{\bm{d}}_i, \hat{\bm{d}}_j, \bm{r}) = -\frac{d_0^2}{4\pi\varepsilon_0 r^3} \left( 3\, \hat{\bm{d}}_i \cdot \hat{\bm{r}}\, \hat{\bm{d}}_j \cdot \hat{\bm{r}} - \hat{\bm{d}}_i \cdot \hat{\bm{d}}_j \right), \quad (5)$$

where $r = |\bm{r}|$ and $\hat{\bm{r}} = \bm{r}/r$. We separate the Hamiltonian into two parts, the non-dipolar part

$$H_{\rm nd} = \sum_{i=1}^{N} \left( \frac{\bm{p}_i^2}{2m} + V_{\rm trap}(\bm{r}_i) \right) + \sum_{i<j} V_s(\bm{r}_i - \bm{r}_j) \quad (6)$$

and the dipolar part

$$H_{\rm d} = \sum_{i=1}^{N} \left( \frac{\bm{L}_i^2}{2I} - d_0 \hat{\bm{d}}_i \cdot \bm{E} \right) + \sum_{i<j} V_{dd}(\hat{\bm{d}}_i, \hat{\bm{d}}_j, \bm{r}_i - \bm{r}_j). \quad (7)$$

Since we consider a Bose-Einstein condensate such that all the bosons are in the same spatial single-particle state $\phi(\bm{r}_i)$, an accurate variational many-body wavefunction is

$$\Psi(\bm{r}_1, \bm{r}_2, \ldots, \bm{r}_N; \bm{d}_1, \bm{d}_2, \ldots, \bm{d}_N) = \psi_d(\bm{d}_1, \bm{d}_2, \ldots, \bm{d}_N) \prod_{i=1}^{N} \phi(\bm{r}_i), \quad (8)$$

---

[*] j.armaitis@uu.nl

where the single-particle wavefunction is normalized as

$$\int d\boldsymbol{r} |\phi(\boldsymbol{r})|^2 = 1. \tag{9}$$

For small electric fields we can in first instance neglect the influence of the dipole-dipole interaction on the spatial wavefunction and $\phi(\boldsymbol{r})$ is determined by minimizing the non-dipolar part of the Hamiltonian. In that case, the energy including the Lagrange multiplier $\mu$ that enforces particle number $N$ is given by

$$\langle H_{\rm nd} \rangle - \mu N = N \int d\boldsymbol{r}\, \phi^*(\boldsymbol{r}) \left( -\frac{\hbar^2 \boldsymbol{\nabla}^2}{2m} + V_{\rm trap}(\boldsymbol{r}) + \frac{2(N-1)\pi\hbar^2 a}{m} |\phi(\boldsymbol{r})|^2 - \mu \right) \phi(\boldsymbol{r}). \tag{10}$$

Variation of $\langle H_{\rm nd} \rangle - \mu N$ with respect to $\phi^*(\boldsymbol{r})$ yields

$$\left( -\frac{\hbar^2 \boldsymbol{\nabla}^2}{2m} + V_{\rm trap}(\boldsymbol{r}) + \frac{4\pi\hbar^2 a}{m}(N-1)|\phi(\boldsymbol{r})|^2 - \mu \right) \phi(\boldsymbol{r}) = 0. \tag{11}$$

For a large cloud ($N \gg 1$), the kinetic energy related to the spatial part of the many-body wavefunction is much smaller than the interaction energy or the trapping potential. Neglecting the kinetic energy is known as the Thomas-Fermi approximation [2], resulting in the wavefunction $\phi \equiv \psi_{\rm TF}$ with

$$|\psi_{\rm TF}(\boldsymbol{r})|^2 = n(\boldsymbol{r})/N = \frac{m}{4\pi\hbar^2 a} \frac{\mu - V_{\rm trap}(\boldsymbol{r})}{N}, \tag{12}$$

where $n(\boldsymbol{r})$ is the density of particles, and we have approximated $N - 1 \simeq N$.

We continue by making a variational ansatz for the dipolar part of the wavefunction. To that end, we construct a superposition of wavepackets of the form

$$\Psi(\boldsymbol{r}_1, \boldsymbol{r}_2, \ldots, \boldsymbol{r}_N; \hat{\boldsymbol{d}}_1, \hat{\boldsymbol{d}}_2, \ldots, \hat{\boldsymbol{d}}_N) = \int d\hat{\boldsymbol{d}} \prod_{i=1}^{N} \left[ \psi_{\rm TF}(\boldsymbol{r}_i) \zeta(\hat{\boldsymbol{d}}_i - \hat{\boldsymbol{d}}) \right] \psi(\hat{\boldsymbol{d}}), \tag{13}$$

where the wavefunction $\zeta$ has the property that individual dipoles point in the $\hat{\boldsymbol{d}}$ direction on average:

$$\int d\hat{\boldsymbol{d}}_i\, |\zeta(\hat{\boldsymbol{d}}_i - \hat{\boldsymbol{d}})|^2 \hat{\boldsymbol{d}}_i = \hat{\boldsymbol{d}}. \tag{14}$$

Moreover, we require $\zeta$ to be sharply localized around $\hat{\boldsymbol{d}}$. Note that $\Psi$ is not a mean-field wavefunction, except for the case of the strong electric field, where $\psi(\hat{\boldsymbol{d}})$ approaches a delta function.

We now come back to the part of $H_{\rm mb}$ that describes the dipolar degree of freedom and consider its expectation value per particle

$$\frac{\langle H_{\rm d} \rangle}{N} = \int \prod_{i=1}^{N} \left[ d\boldsymbol{r}_i\, d\hat{\boldsymbol{d}}_i \right] \Psi^*(\boldsymbol{r}_1, \boldsymbol{r}_2, \ldots, \boldsymbol{r}_N; \hat{\boldsymbol{d}}_1, \hat{\boldsymbol{d}}_2, \ldots, \hat{\boldsymbol{d}}_N) H_{\rm d} \Psi(\boldsymbol{r}_1, \boldsymbol{r}_2, \ldots, \boldsymbol{r}_N; \hat{\boldsymbol{d}}_1, \hat{\boldsymbol{d}}_2, \ldots, \hat{\boldsymbol{d}}_N) \tag{15}$$

in order to derive the Hamiltonian for the direction of the macroscopic dipole moment $\boldsymbol{d}$. In what follows, we consider this expectation value term by term. We start with the rotation energy for a molecule, i.e.,

$$\int \prod_{i=1}^{N} \left[ d\boldsymbol{r}_i\, d\hat{\boldsymbol{d}}_i \right] \Psi^* \frac{\boldsymbol{L}_j^2}{2I} \Psi = \frac{-\hbar^2}{2I} \int \prod_{i=1}^{N} \left[ d\boldsymbol{r}_i\, d\hat{\boldsymbol{d}}_i \right] \Psi^* \left( \hat{\boldsymbol{d}}_j \times \frac{\partial}{\partial \hat{\boldsymbol{d}}_j} \right)^2 \Psi \tag{16}$$

$$\simeq \frac{-\hbar^2}{2I} \int d\hat{\boldsymbol{d}}\, d\hat{\boldsymbol{d}}'\, \psi^*(\hat{\boldsymbol{d}}') \left( \hat{\boldsymbol{d}} \times \frac{\partial}{\partial \hat{\boldsymbol{d}}} \right)^2 \delta(\hat{\boldsymbol{d}} - \hat{\boldsymbol{d}}') \psi(\hat{\boldsymbol{d}}) = \int d\hat{\boldsymbol{d}}\, \psi^*(\hat{\boldsymbol{d}}) \frac{\boldsymbol{L}^2}{2I} \psi(\hat{\boldsymbol{d}}), \tag{17}$$

where $\boldsymbol{L}$ is the angular momentum operator for the average direction $\hat{\boldsymbol{d}}$ and we have used the localization property

$$\prod_{i=1}^{N} \left( \int d\hat{\boldsymbol{d}}_i\, \zeta^*(\hat{\boldsymbol{d}}_i - \hat{\boldsymbol{d}}') \zeta(\hat{\boldsymbol{d}}_i - \hat{\boldsymbol{d}}) \right) \simeq \delta(\hat{\boldsymbol{d}} - \hat{\boldsymbol{d}}'). \tag{18}$$

Since the electric field term follows from Eq. (14) in a straightforward manner, we are left with the dipole-dipole interaction term. Similarly to the wavefunction, it can be written as a product of the spatial part and the dipolar part:

$$V_{dd}(\hat{\boldsymbol{d}}_i, \hat{\boldsymbol{d}}_j, \boldsymbol{r}_i - \boldsymbol{r}_j) = \hat{d}_i^\alpha A^{\alpha\beta}(\boldsymbol{r}_i - \boldsymbol{r}_j)\hat{d}_j^\beta, \tag{19}$$

where $\alpha$ and $\beta$ are the spatial indices and a sum over them is implied. We now rewrite the dipole-dipole interaction expectation value making the average dipole direction dependence explicit and having in mind that $i \neq j$:

$$\int \prod_{k=1}^N \left[ \mathrm{d}\boldsymbol{r}_k \, \mathrm{d}\hat{\boldsymbol{d}}_k \right] \mathrm{d}\hat{\boldsymbol{d}} \, \mathrm{d}\hat{\boldsymbol{d}}' \, \Psi^* V_{dd}(\hat{\boldsymbol{d}}_i, \hat{\boldsymbol{d}}_j, \boldsymbol{r}_i - \boldsymbol{r}_j)\Psi$$

$$= \frac{1}{N^2} \int \mathrm{d}\boldsymbol{r}_i \, \mathrm{d}\boldsymbol{r}_j n(\boldsymbol{r}_i)n(\boldsymbol{r}_j)A^{\alpha\beta}(\boldsymbol{r}_i - \boldsymbol{r}_j) \int \mathrm{d}\hat{\boldsymbol{d}} \, \mathrm{d}\hat{\boldsymbol{d}}' \prod_{k=1}^N \left[\mathrm{d}\hat{\boldsymbol{d}}_k \zeta^*(\hat{\boldsymbol{d}}_k - \hat{\boldsymbol{d}}')\zeta(\hat{\boldsymbol{d}}_k - \hat{\boldsymbol{d}})\right] \hat{d}_i^\alpha \hat{d}_j^\beta \psi^*(\hat{\boldsymbol{d}}')\psi(\hat{\boldsymbol{d}}) \tag{20}$$

$$= \frac{1}{N^2} \int \mathrm{d}\boldsymbol{r}_i \, \mathrm{d}\boldsymbol{r}_j n(\boldsymbol{r}_i)n(\boldsymbol{r}_j)A^{\alpha\beta}(\boldsymbol{r}_i - \boldsymbol{r}_j) \int \mathrm{d}\hat{\boldsymbol{d}} \, \hat{d}^\alpha \hat{d}^\beta |\psi(\hat{\boldsymbol{d}})|^2, \tag{21}$$

where we have used the localization property again. Using the definition of the probability to find two particles a distance $\boldsymbol{r}$ apart

$$P(\boldsymbol{r}) = \frac{1}{N^2} \int \mathrm{d}\boldsymbol{x} \, n(\boldsymbol{x})n(\boldsymbol{x} + \boldsymbol{r}) \tag{22}$$

and performing a sum over pairs of particles yields the following dipole-dipole interaction energy per particle:

$$V_{\mathrm{dd}}^{\mathrm{TF}} = -\frac{(N-1)d_0^2}{4\pi\varepsilon_0} \int \mathrm{d}\boldsymbol{r} P(\boldsymbol{r})\frac{1}{r^5}\left(3\,(\hat{\boldsymbol{d}} \cdot \boldsymbol{r})^2 - \hat{\boldsymbol{d}}^2 r^2\right). \tag{23}$$

Since $P(-\boldsymbol{r}) = P(\boldsymbol{r})$, we rewrite the last expression as

$$V_{\mathrm{dd}}^{\mathrm{TF}} = -\frac{(N-1)d_0^2}{4\pi\varepsilon_0} \int \mathrm{d}\boldsymbol{r} P(\boldsymbol{r})\frac{1}{r^5}\hat{d}^\alpha \left(3 r^\alpha r^\beta - r^2 \delta^{\alpha\beta}\right)\hat{d}^\beta. \tag{24}$$

We find that only the diagonal terms survive the integration and after going to cylindrical coordinates we find that

$$\int \mathrm{d}\boldsymbol{r} P(\boldsymbol{r})\frac{1}{|r|^5}\left(3(r^x)^2 - r^2\right) = 2\pi \int \rho\,\mathrm{d}\rho\,\mathrm{d}z P(\rho,z)\frac{1}{r^3}\left[\frac{3}{2}\frac{\rho^2}{r^2} - 1\right] \tag{25}$$

and

$$\int \mathrm{d}\boldsymbol{r} P(\boldsymbol{r})\frac{1}{|r|^5}\left(3(r^z)^2 - r^2\right) = 2\pi \int \rho\,\mathrm{d}\rho\,\mathrm{d}z P(\rho,z)\frac{1}{r^3}\left[-3\frac{\rho^2}{r^2} + 2\right]. \tag{26}$$

Since $(\hat{d}^x)^2 + (\hat{d}^y)^2 - 2(\hat{d}^z)^2 = (\hat{\boldsymbol{d}})^2 - 3(\hat{d}^z)^2$, and by approximating $N-1 \simeq N$ again we arrive at the final expression for the dipole-dipole interaction

$$V_{\mathrm{dd}}^{\mathrm{TF}} = -\frac{Nd_0^2}{4\pi\varepsilon_0} \int \mathrm{d}\boldsymbol{r} P(\boldsymbol{r})\frac{1}{r^5}\left(3\,(\hat{\boldsymbol{d}} \cdot \boldsymbol{r})^2 - \hat{\boldsymbol{d}}^2 r^2\right) = C_{\mathrm{dd}}(3\hat{d}_z^2 - \hat{\boldsymbol{d}}^2), \tag{27}$$

where $C_{\mathrm{dd}}$ is the effective dipolar interaction strength

$$C_{\mathrm{dd}} = \frac{d_0^2 N}{4\varepsilon_0} \int \mathrm{d}z\rho\,\mathrm{d}\rho P(R)\frac{1}{r^3}\left(\frac{3}{2}\frac{\rho^2}{r^2} - 1\right). \tag{28}$$

Hence, the expectation value of the dipolar part of the many-body Hamiltonian is

$$\frac{\langle H_{\mathrm{d}} \rangle}{N} = \int \mathrm{d}\hat{\boldsymbol{d}}\,\psi^*(\hat{\boldsymbol{d}})\left(\frac{\boldsymbol{L}^2}{2I} - d_0\hat{\boldsymbol{d}} \cdot \boldsymbol{E} + C_{\mathrm{dd}}(3\hat{d}_z^2 - \hat{\boldsymbol{d}}^2)\right)\psi(\hat{\boldsymbol{d}}). \tag{29}$$

Finally, we vary this expression with respect to $\psi^*(\hat{\boldsymbol{d}})$ and obtain the single-mode approximation Hamiltonian given in the Letter

$$H = \frac{\boldsymbol{L}^2}{2I} - d_0\hat{\boldsymbol{d}} \cdot \boldsymbol{E} + C_{\mathrm{dd}}(3\hat{d}_z^2 - \hat{\boldsymbol{d}}^2), \tag{30}$$

where this last Hamiltonian is later used to determine the ground-state wavefunction $\psi(\hat{\boldsymbol{d}})$.

## II. PROBABILITY DISTRIBUTION

We would like to point out that in order to draw any conclusions about the state of the system by comparing $P_{l,m}$ measurements, where $P_{l,m}$ is the probability to occupy a state with total angular momentum $l$ and its projection $m$, it is important to diagonalize the $\langle \hat{d}_i \hat{d}_j \rangle$ matrix for each configuration of the electric field $\boldsymbol{E}$ and the dipolar interaction strength $C_{\text{dd}}$. To illustrate this point, we provide $P_{1,m}$ for two different situations: a "squeezed" state with both a finite electric field and a finite dipole-dipole interaction strength (Fig. 1 left) and a dipolar state, where the dipole-dipole interaction is zero (Fig. 1 center and right). The two situations give a qualitatively similar $P_{1,m}$ when viewed in the same coordinate system (Fig. 1 left and center). However, if one compares the probability distributions for the coordinate systems where $\langle \hat{d}_i \hat{d}_j \rangle$ is diagonal (Fig. 1 left and right), the difference is obvious.

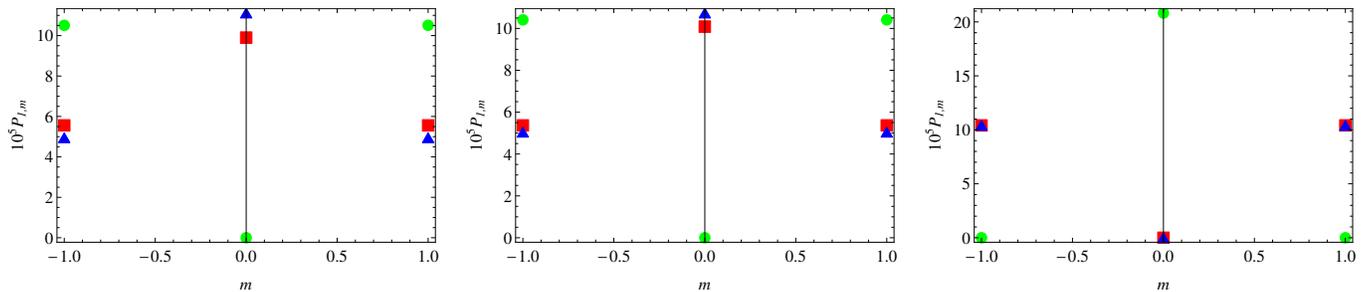

FIG. 1. Probability $P_{1,m}$ of occupying a state with total angular momentum 1 and its projection $m$ for the electric field $E = 0.05 \hbar^2 / 2 I d_0$ ($\boldsymbol{E}$ is at $\pi/4$ angle to the $z$ axis). The left figure depicts a situation with a finite dipole-dipole interaction $C_{\text{dd}} = 0.1 \hbar^2 / 2I$, whereas the other figures are for the non-interacting ($C_{\text{dd}} = 0$) case. The red squares correspond to the $x'$ direction, the green circles correspond to the $y'$ direction, and the blue triangles correspond to the $z'$ direction. The axes are the same for the left and center figures, while they are different for the right figure. They are defined such that the $\langle \hat{d}_i \hat{d}_j \rangle$ matrix is diagonal and has its smallest eigenvalue in the $z'$ direction for the $C_{\text{dd}} = 0.1 \hbar^2 / 2I$ case (left and center) or for $C_{\text{dd}} = 0$ (right).



\end{document}